\documentclass[pre,twocolumn]{revtex4-1}
\usepackage{bm}
\usepackage{graphicx}
\usepackage{amsmath}
\usepackage{amscd}
\usepackage{epstopdf}
\usepackage{verbatim}
\usepackage[usenames]{color}
\usepackage{array}
\usepackage{xspace}
\usepackage{xstring}
\usepackage[normalem]{ulem}

\newcommand{\ie}{\textit{i.e.}\xspace}

\newcommand{\Dr}{D_\text{r}}  

\definecolor{Blue}{rgb}{0,0,0.8}
\definecolor{Red}{rgb}{0.8,0,0}
\definecolor{Green}{rgb}{0,0.5,0}
\definecolor{Purple}{rgb}{1,0,1}

\newcommand\yfsout{\bgroup\markoverwith{\textcolor{Red}{\rule[.5ex]{2pt}{0.4pt}}}\ULon}

\renewcommand{\vec}[1]{{\bf #1}}

\newcommand{\vecug}[1]{\hat{\bm #1}}

\begin{document}

\title{Dynamics of Self-Propelled Particles Under Strong Confinement}
\author{Yaouen Fily, Aparna Baskaran, Michael F. Hagan}
\affiliation{Martin Fisher School of Physics, Brandeis University, Waltham, MA 02453, USA}
\date{\today}


\begin{abstract}
We develop a statistical theory for the dynamics of non-aligning, non-interacting self-propelled particles confined in a convex box in two dimensions.
We find that when the size of the box is small compared to the persistence length of a particle's trajectory (strong confinement), the steady-state density is zero in the bulk and proportional to the local curvature on the boundary.
Conversely, the theory may be used to construct the box shape that yields any desired density distribution on the boundary.
When the curvature variations are small, we also predict the distribution of orientations at the boundary and the exponential decay of pressure as a function of box size recently observed in 3D simulations in a spherical box. \end{abstract}

\maketitle

Active fluids consisting of self-propelled units are found in biology on scales ranging from the dynamically reconfigurable cell cytoskeleton~\cite{Schaller2011} to swarming bacterial colonies~\cite{Dombrowski2004,Peruani2012}, healing tissues~\cite{Poujade2007,Trepat2009}, and flocking animals~\cite{Ballerini2008}. Experiments have begun to achieve the extraordinary capabilities and emergent properties of these biological systems in nonliving active fluids of self-propelled particles, consisting of chemically~\cite{Palacci2010a,Paxton2004, Hong2007,Jiang2010,Volpe2011,Thutupalli2011} or electrically~\cite{Bricard2013} propelled colloids, or monolayers of vibrated granular particles \cite{Narayan2007,Kudrolli2008,Deseigne2010}.

In contrast to thermal motion, active motion is correlated over experimentally accessible time and length scales. When the persistence length of active motion becomes comparable to the mean free path, uniquely active effects arise that transcend the thermodynamically allowed behaviors of equilibrium systems, including giant number fluctuations and spontaneous flow~\cite{Vicsek1995,Toner1995,Toner2005,Ramaswamy2003,Mishra2006,Peruani2006,Narayan2007,Deseigne2010,Yang2010,Peruani2011a,Peruani2012, Chate2006,Chate2008,Peruani2011,Peruani2013,Ramaswamy2010,Marchetti2013}.
Importantly, a sufficient active persistence length is the only requirement for macroscopic manifestations of activity, as revealed by athermal phase separation of non-aligning, repulsive self-propelled particles~\cite{Tailleur2008,Thompson2011,Fily2012,Redner2013, Bialke2013,Cates2013,Stenhammar2013,Redner2013b,Fily2013,Stenhammar2014,Wittkowski2014}.

When boundaries and obstacles are patterned on the scale of the active correlation length, they dramatically alter the dynamics of the system, and striking macroscopic properties emerge~\cite{Wan2008,Tailleur2009,Angelani2011,Ghosh2013,Ai2013,Nash2010,Reichhardt2013,Reichhardt2014};
for example, ratchets and funnels drive spontaneous flow in active fluids~\cite{Wan2008,Tailleur2009,Angelani2011,Ghosh2013,Ai2013}. This effect has been used to direct bacterial motion~\cite{Galajda2007} and harness bacterial power to propel microscopic gears~\cite{Angelani2009,DiLeonardo2010,Sokolov2010}.
However, optimizing such devices for technological applications requires understanding the interaction of an active fluid with boundaries of arbitrary shape. More generally, any real-world device necessarily includes boundaries, and thus the effects of boundary size and shape are essential design parameters.
Although recent studies have explored confinement in simple geometries~\cite{Tailleur2009,Nash2010,Elgeti2013,Lee2013,Mallory2013}, there is no general theory for the effect of boundary shape.

\begin{figure}[h]
\centering
\includegraphics[width=0.95\linewidth]{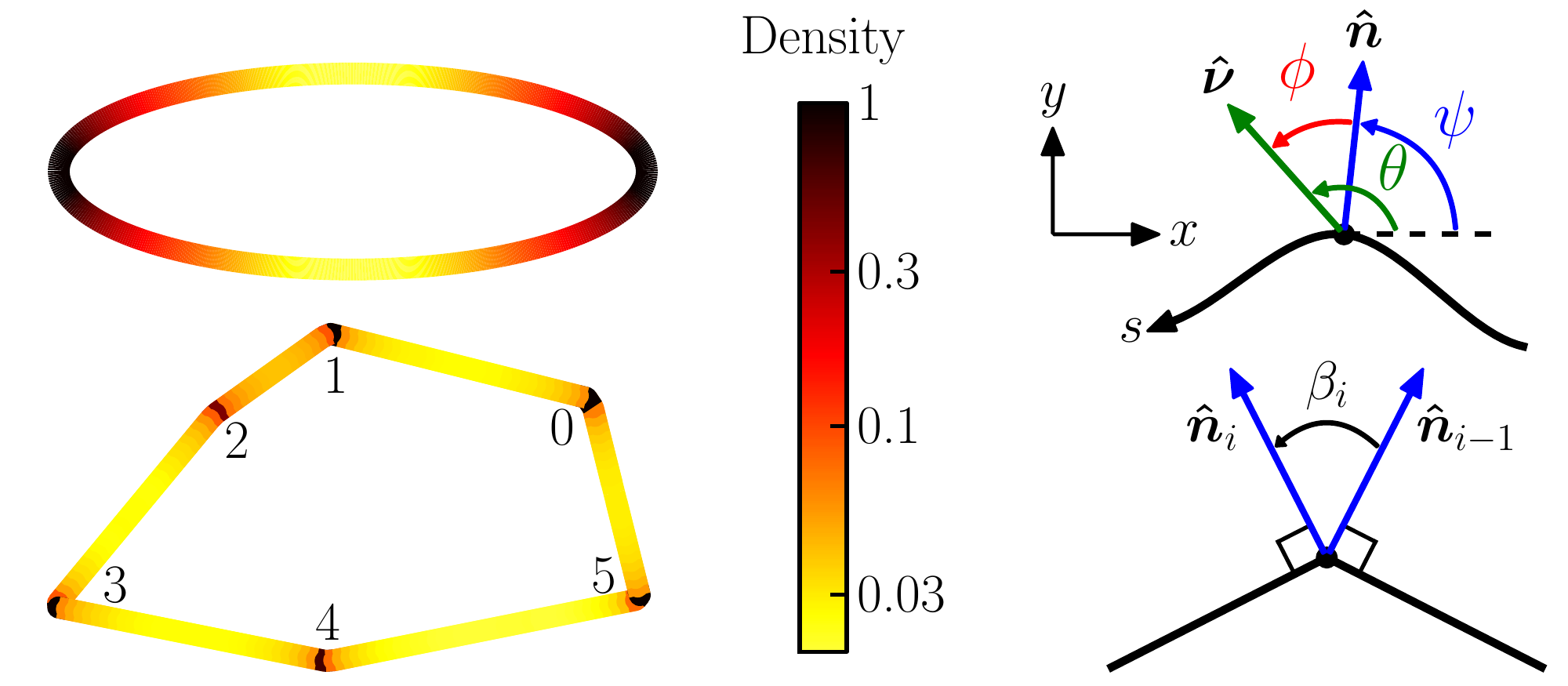} 
\caption{
Left:
Visual summary of simulation results showing particles concentrated in high curvature regions.
Right:
Geometric notations for the analytic theory for a smooth boundary (top) and a polygon (bottom).
}
\label{fig:sketches}
\end{figure}

In this Letter, we study the dynamics of non-aligning and non-interacting self-propelled particles confined to two-dimensional convex containers, such as ellipses and polygons. We find that the boundary shape dramatically affects the active fluid's dynamics and thermomechanical properties in the limit of ``strong confinement'', in which the container size is small compared to the active persistence length (the distance a particle travels before its orientation decorrelates).
In particular: (i) particles are confined to the boundary, (ii) the steady-state distribution of particles at the boundary is proportional to the local curvature (see Fig.~\ref{fig:sketches}), and
(iii) when the curvature varies slowly, the local pressure exerted on the boundary decays exponentially with the ratio of the radius of curvature to the active persistence length.
Results (i) and (ii) are derived in the limit of small and slowly varying curvature radius, then extended to polygonal boxes.
They likely hold for arbitrary convex boundaries, although the definition of ``strong confinement'' depends on the type of boundary.
Result (iii) explains recent pressure measurements on 3D active particles in spherical confinement~\cite{Mallory2013}.


\emph{Model} ---
We consider an overdamped self-propelled particle with position ${\bf r}$ and orientation $\hat{\bm \nu}=\cos\theta\,\hat{\bf x} + \sin\theta\,\hat{\bf y}$ whose dynamics is described by
\begin{align}
\dot{\vec{r}} = v_0 \vecug{\nu} + \mu \vec{F}_\text{wall}
\, , \quad
\dot{\theta} = \xi(t)
\label{eq:eom1}
\end{align}
where $v_0$ is the self-propulsion speed, $\mu$ is the mobility, $\xi$ is a white Gaussian noise with zero mean and correlations $\langle\xi(t)\xi(t')\rangle=2\Dr\delta(t-t')$, and over-dots indicate time derivatives.
The hard wall exerts a force $\vec{F}_\text{wall}=-v_0(\vecug{\nu}\cdot\hat{\bf n})\hat{\bf n}/\mu$ if the particle is at the wall and $\vecug{\nu}\cdot\hat{\bf n}>0$ and zero otherwise, where
$\hat{\bf n}=\cos\psi\,\hat{\bf x} + \sin\psi\,\hat{\bf y}$ is the local normal to the wall pointing outwards; i.e., the normal component of the velocity that would drive the particle into the wall is cancelled by the wall force.

When the particle is at the wall, its configuration is characterized by its arclength $s\in [0,L)$ along the boundary, where $L$ is the box perimeter, and its orientation $\phi=\theta-\psi$ relative to the local boundary normal (see Fig.~\ref{fig:sketches}).
Projecting Eq.~\eqref{eq:eom1} onto the boundary tangent yields equations of motion for $s$ and $\phi$:
\begin{align}
\dot{s} = v_0 \sin\phi
\, , \quad
\dot{\phi} = \xi(t) - \frac{v_0}{R}\sin\phi
\label{eq:eom2}
\end{align}
where $R(s)$ is the local radius of curvature, which satisfies $\dot{\psi}=\dot{s}/R$.
We only consider convex boxes for which $R>0$; thus $\phi=0$ is a stable equilibrium point with characteristic relaxation time $R/v_0$.
The corresponding restoring force
acts by moving the particle along the boundary until the wall's normal aligns with its orientation.

We now argue that
in the limit of ``strong confinement'', $\phi$ is small. As a result, the particle never leaves the boundary (this would require $|\phi|>\pi/2$) and its dynamics is entirely described by Eqs.~\eqref{eq:eom2}, which we may linearize about $\phi=0$~%
\footnote{After linearization, Eqs.~\eqref{eq:eom2} become a Langevin equation for a free 1D particle of position $s$ and velocity $v_0\phi$ with position-dependent friction and temperature.}%
.
%
In the case of a circle (constant $R$), the resulting equation of motion for $\phi$ can be directly integrated:
%
\begin{align}
\phi(t) = \int_0^t dt'\, \xi(t') e^{-v_0(t-t')/R}
\label{eq:formal_phi}.
\end{align}
It follows that $\phi$ is a Gaussian random variable with zero mean and $\langle\phi^2\rangle=R\Dr/v_0$.
Physically, $\phi$ is small when the curvature radius $R$ is much smaller than the particle's persistence length $v_0/\Dr$; we refer to this case as the strong confinement limit.
In the rest of this Letter we consider the implications of this result for the particle density and its extension to arbitrary convex boxes.


\smallskip
\emph{Statistical Description} ---
Let $f(s,\phi,t)$ be the probability density of finding a particle with relative orientation $\phi$ at arclength $s$ at time $t$. After linearizing \eqref{eq:eom2}, $f$ obeys the Smoluchowski equation:
\begin{align}
\label{eq:smol}
\partial_t f = - v_0\, \phi\, \partial_s f + \frac{v_0}{R} f
    + \frac{v_0\, \phi}{R} \partial_\phi f + \Dr \partial_{\phi^2} f
\end{align}
The boundary is closed and particles leave the boundary when $|\phi|>\pi/2$;  therefore physical solutions satisfy $f(s+L,\phi,t)=f(s,\phi,t)$ and $f=0$ for $|\phi| \ge \pi/2$.
Since we work in the small $\phi$ limit, we  assume
$f(s,\pm\pi/2,t)=\partial_\phi f(s,\pm\pi/2,t)=0$~\footnote{%
Since the Smoluchowski equation is second order in $\phi$,  these boundary conditions are over-constraining. Physically, however, no particle ever gets close to $\phi=\pm\pi/2$ and thus the boundary terms are irrelevant.}%
, and seek  the steady state solution to \eqref{eq:smol}.

To this end, we introduce the moments $g_n(s)=\int d\phi\,\phi^n f(s,\phi)$ such that
$\rho=g_0$ is the density of particles at the boundary and $\langle\phi^n\rangle=g_n/g_0$.
The steady state solution to \eqref{eq:smol} is then obtained by solving the recurrence relation  $\partial_s g_{n+1} + \frac{n}{R} g_n - n(n-1)\frac{\Dr}{v_0} g_{n-2}= 0$, the first three equations of which are:
\begin{align}
\label{eq:moment12}
\partial_s g_1 = 0
\, , \quad
\partial_s g_2 + \frac{1}{R} g_1 = 0
\\
\label{eq:moment3}
\partial_s g_3 + \frac{2}{R} g_2 - \frac{2\Dr}{v_0} g_0 = 0
\end{align}
From Eqs.~\eqref{eq:moment12}
it follows that $g_1$ is constant and
$g_2(s) = g_2(0)-g_1 \int_0^s du/R(u)$.
Like $f$, $g_2$ is a periodic function of $s$, and $\oint_0^L du/R(u)=2\pi$ for any planar curve~\cite{Kamien2002};
therefore $g_1$ must be zero, i.e. there is no density flux at steady state.
We close the system by neglecting $\partial_s g_3$.
The approximation is exact for circular boxes for which $\phi$'s gaussianity implies $g_3=0$, and should hold when $R$ is nearly constant.
It can also be interpreted as setting the third cumulant to zero; a standard closure method.
Finally, the normalization constraint $\int_0^L ds\,\rho(s)=N$ with $N$ as the total number of particles, gives
\begin{align}
\label{eq:density_variance}
\rho (s) = \frac{N}{2\pi R}
\end{align}
and $\langle \phi^2(s) \rangle = R \Dr/v_0$. 
Eq.~\eqref{eq:density_variance} is our primary result. The density of particles at the boundary is inversely proportional to the local curvature radius; \ie, regions of high curvature act as attractors for active particles.
A more general derivation of this result can be found in appendix~\ref{ap:overdamped}.
The second key result is that fluctuations in $\phi$ are controlled by $R\Dr/v_0$, consistent with the premise that $\phi$ is small under strong confinement. 
This result is limited by the validity of our closure approximation and its scope and relevance are discussed  below.


\smallskip
\emph{Pressure} ---
When $\phi$ is nearly gaussian, we can compute the local pressure
exerted on the boundary,
\begin{align}
P(s)
= \frac{v_0}{\mu} \rho\, \langle \cos\phi \rangle
= \frac{N\,v_0}{2\pi\mu R}\, e^{-\frac{R \Dr}{2v_0}}
\label{eq:pressure}
\end{align}
where $v_0/\mu$ is the force exerted by a single particle aligned with the normal and
we  used $\langle\cos\phi\rangle=e^{-\langle\phi^2\rangle/2}$.
The exponential decay of pressure with $R \Dr/v_0$ was recently observed in 3D simulations of active particles in a spherical box~\cite{Mallory2013}.
Although curvature on a surface is a tensor, for a sphere it reduces to a scalar and we expect the same exponential dependence on $R$ as shown here for 2D, though with a different numerical factor.
Also, in the limit $\phi\ll1$, the exponential tends to unity and pressure is essentially proportional to curvature.
Expanding about $\phi=0$ and using $\langle\phi\rangle=g_1/g_0=0$, we may write $\langle e^{\phi}\rangle\approx 1-\langle\phi^2\rangle/2$; therefore $P\propto e^{-\langle\phi^2\rangle/2}$ remains correct asymptotically, independent of the distribution of $\phi$.


\begin{figure}[h]
\centering
\includegraphics[width=\linewidth]{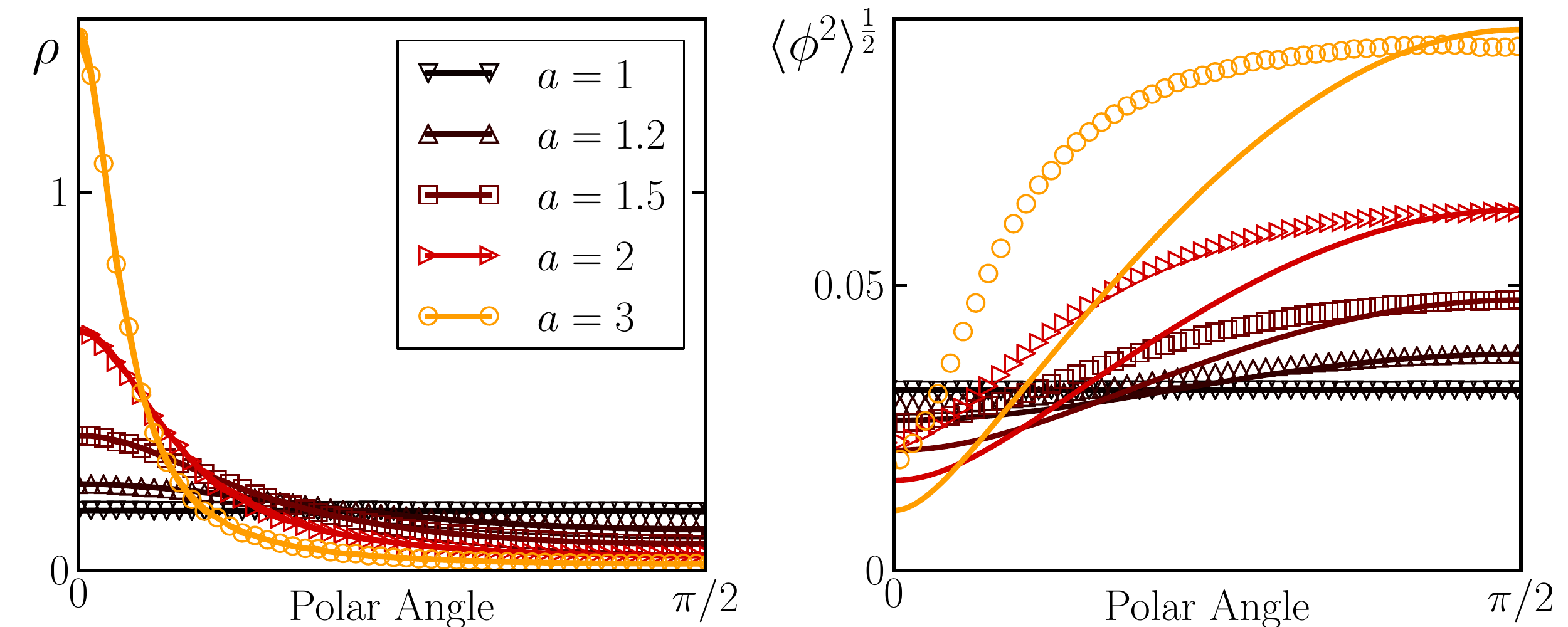} 
\caption{
Boundary density $\rho$ (left) and standard deviation $\langle\phi^2\rangle^{1/2}$ of the orientation relative to the boundary normal (right) as a function of polar angle in the first quadrant of an elliptic box with semi-axes $a$ and $1$ aligned with the $x$ and $y$ axes respectively.
Circles are from simulations with $\Dr=10^{-3}$. Solid lines are from Eqs.~\eqref{eq:density_variance}.
Both $\rho$ and $\langle\phi^2\rangle$ are symmetric with respect to reflections about $0$ and $\pi/2$.
}
\label{fig:ellipse_rho_sigma}
\end{figure}

\smallskip
\emph{Simulations} ---
%
To explore the domain of validity of our statistical theory and 
the physics beyond
the low moment closure, we performed molecular dynamics simulations of Eq.~\eqref{eq:eom1}.
We consider $v_0=1$ and various $\Dr$ in elliptical boxes with semi-axes $a>1$ and $1$ aligned with the $x$ and $y$ axes respectively.
We plot results against the polar angle $\alpha$ rather than the arclength; thus the curvature radius oscillates between $R=a^2$ at $\alpha=0,\,\pi$ and $R=a^{-1}$ at $\alpha=\pm\pi/2$.

The simulation results are shown in Fig.~\ref{fig:ellipse_rho_sigma}.
As expected, in the circular case ($a=1$) the distribution of $\phi$ (not shown) is gaussian and both $\rho$ and $\langle\phi^2\rangle$ match the theory perfectly.
Near-perfect agreement between Eq.~\eqref{eq:density_variance} and the observed density $\rho$ persists at all simulated aspect ratios.
The magnitude and qualitative behavior of $\langle\phi^2\rangle$ remain well captured as well, but quantitative agreement is lost with increasing $a$, even though $R \Dr/v_0$ is small.
This results from the breakdown of the $\partial_s g_3=0$ assumption as the distribution of $\phi$ departs from gaussianity (see appendix~\ref{ap:var_skew_kurt}). 
To improve the theory, one may push the moment closure to higher orders; i.e., set the $n^\text{th}$ cumulant to zero for some $n>3$. 
This leads to a non-linear equation for $\rho$ and $g_2$ that involves derivatives of $\rho$ and $R$ with respect to $s$, even for $n=4$. These terms suggest that the prediction
$\langle \phi^2 \rangle=R\Dr/v_0$
requires not only $R \Dr/v_0\ll1$ but also $dR/ds\ll1$, and thus may only hold in slightly deformed circular boxes.

To understand why the low moment closure successfully  predicts $\rho$ even when it poorly describes $\langle\phi^2\rangle$, we consider the limit case $\phi=0$, or $\theta=\psi$,  in which a particle is always located at the position $s$ where its orientation aligns with the boundary normal.
Since the dynamics of $\theta$ is purely diffusive, its steady-state distribution is flat: $\rho(\theta)=\rho(\psi)=N/(2\pi)$.
A change of variable then yields $\rho(s)=(d\psi/ds)\rho(\psi)\propto1/R$.
In other words, for sufficiently small fluctuations of $\phi=\theta-\psi$,
$\rho(s)$ is controlled by $d\psi/ds$ and is essentially independent of the form of the distribution of $\phi$.
This reasoning only requires $\psi(s)$ to be monotonic and should apply to any convex box.


\smallskip	
\emph{Polygonal Boxes} ---
The previous paragraph suggests that Eq.~\eqref{eq:density_variance} applies to arbitrary convex boxes with no restriction on the magnitude of curvature, provided $\phi$ is small.  However, under what conditions is $\phi$ small in such a container, and do these conditions correspond to the strong confinement limit defined above?
To elucidate this point, we turn to a class of shapes for which both $R$ and $dR/ds$ are unbounded, namely polygons.

The radius of curvature is now discontinuous, equal to zero along the edges and infinity at the corners. $\psi(s)$ is a step function with value $\psi_i$ on the edge connecting corners $i$ and $i-1$ (corner indices are defined modulo the number of corners).
The dynamics at corners follows from that at edges: a particle leaves corner $i$ along edge $i$ (resp. edge $i-1$) as soon as $\theta>\psi_i$ (resp. $\theta<\psi_{i-1}$). Conversely, a particle remains stuck at corner $i$ as long as $\theta\in[\psi_{i-1},\psi_i]$.
The numerical results shown in this section were obtained with the polygonal box pictured on Fig.~\ref{fig:sketches}, which has a wide variety of angles and a perimeter $L\approx8.93$ (\ie a radius of order one). Similar results were obtained with different boxes.

\begin{figure}[b]
\centering
\includegraphics[width=\linewidth]{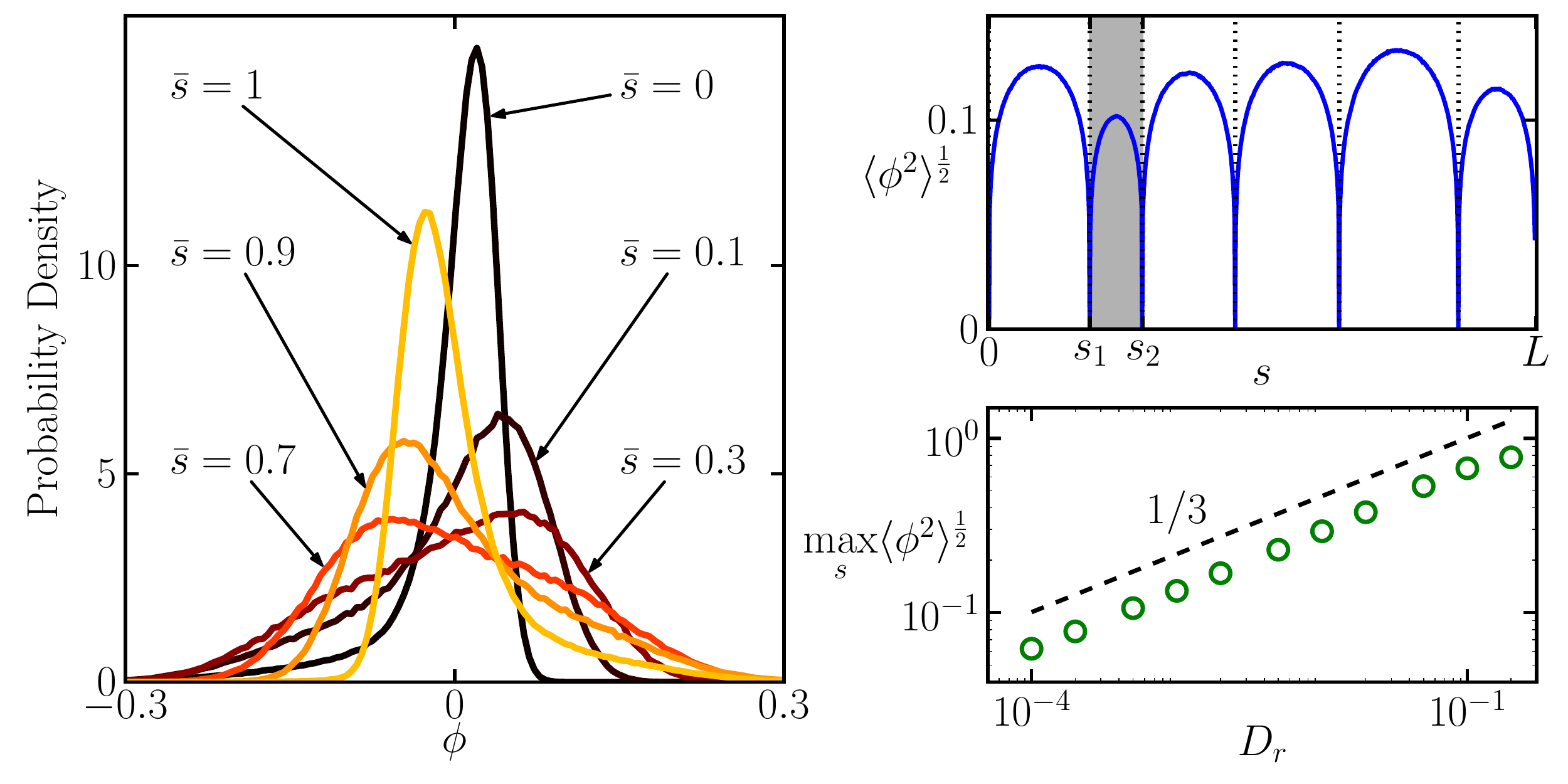} 
\caption{
Left:
Distribution of the orientation $\phi$ relative to the boundary normal at several normalized arclengths $\bar{s}=(s-s_1)/(s_2-s_1)$ between corners $1$ and $2$ ($\bar{s}=0$ at corner $1$, $\bar{s}=1$ at corner $2$) of the polygon shown on Fig.~\ref{fig:sketches} for $\Dr=10^{-3}$.
Top right:
Standard deviation of $\phi$ as a function of arclength for $\Dr=10^{-3}$.
The dotted lines indicate the positions of the corners.
The gray area represents the region between corners $1$ and $2$ from which the distributions of the left panel are extracted.
Bottom right:
Upper bound of the standard deviation of $\phi$ across the boundary as a function of $\Dr$.
The dashed line is a power law with slope $1/3$ as suggested by the analysis in the text.
}
\label{fig:polygon_phi_sigma}
\end{figure}

A particle stuck at a corner automatically satisfies $\phi=0$, since $\phi$ is the angle between the particle's active force and the wall's reaction, which are equal and opposite when the particle is not moving.
%
The case of edges is trickier.
The left panel of Fig.~\ref{fig:polygon_phi_sigma} shows the distribution of $\phi$ at several points along an edge.
It is heavily skewed and
the prediction $\langle\phi^2\rangle=R\Dr/v_0=\infty$ is clearly wrong.
More importantly, the distribution 
is non-local: it changes along the edge, despite every interior point having the same local geometry.
The linearized equation of motion on the edge is $\ddot{s}=v_0\xi(t)$, known as a ``randomly accelerated process''.
Despite its apparent simplicity, the boundary conditions at the corners make its solution difficult \footnote{The case of absorbing boundaries is discussed in the context of first passage processes in Refs.~\cite{Masoliver1995,Redner2001}.}.
However, the scaling of the angular spread $\langle\phi^2\rangle^{1/2}$ can be obtained as follows.
For a particle that escapes a corner with initial angle $\phi=0$, the root mean square velocity and position grow as $v=v_0\phi\sim v_0 (\Dr t)^{1/2}$ and $s \sim \int_0^t dt' v(t') \sim v_0 \Dr^{1/2} t^{3/2}$, respectively, until the particle
reaches a corner again (either its starting point or a neighboring corner).
%
The typical time required to cross an edge of length $\ell$ is obtained by inverting $s(t)=\ell$, leading to $t\sim \ell^{2/3}/(v_0^{2/3}\Dr^{1/3})$~\footnote{The same scaling is found for the mean escape time with absorbing boundaries~\cite{Masoliver1995}.}.
The typical angle reached in the process is then $\phi\sim (\Dr t)^{1/2} \sim (\ell \Dr/v_0)^{1/3}$, consistent with our numerical observations (see bottom right panel of Fig.~\ref{fig:polygon_phi_sigma}).
Most importantly, the angular spread $\langle\phi^2\rangle^{1/2}$ goes to zero in the small box limit despite the infinite radius of curvature,
albeit with a slower decay than previously. 
%

\begin{figure}[t]
\centering
\includegraphics[width=0.9\linewidth]{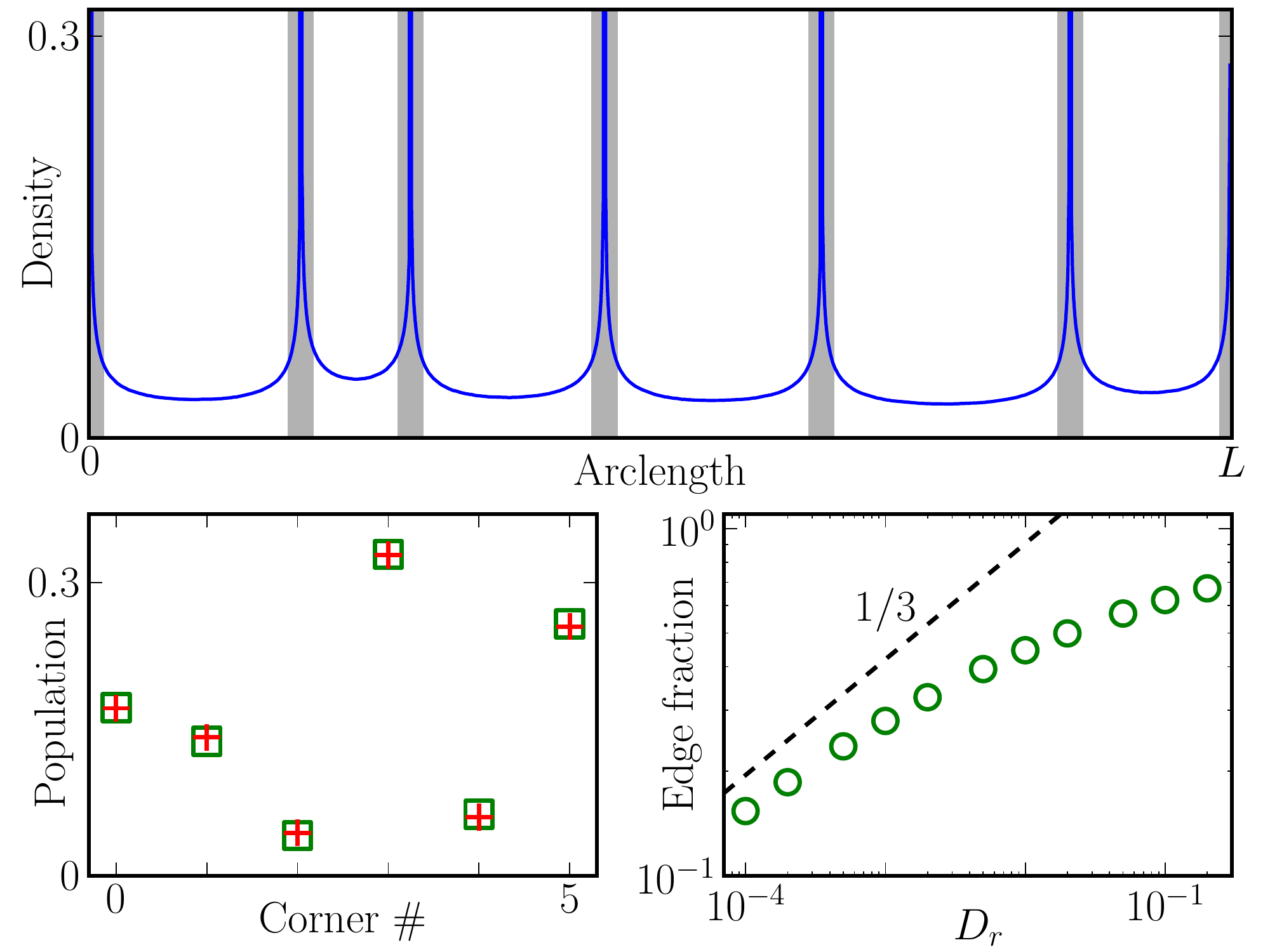} 
\caption{
Top:
Boundary density $\rho$ as a function of arclength for the polygonal box of Fig.~\ref{fig:sketches} with perimeter $L\approx8.93$.
Bottom left:
Observed (crosses) and predicted (squares) corner populations. The former are obtained by integrating the density over each grey region of the top panel. The latter are renormalized to account for edge particles; i.e., particles not in any of the grey regions.
Bottom right:
Fraction of edge particles as a function of the angular noise $\Dr$.
A power law with slope $1/3$ is shown for reference.
The edge fraction for the other two panels ($\Dr=10^{-3}$) is $28\%$.
}
\label{fig:polygon_rho}
\end{figure}

Thus, we may approximate the density by
\begin{align}
\rho(s)= \frac{d\psi}{ds} \rho(\psi) = \frac{N}{2\pi} \sum_i 
\beta_i \delta(s-s_i)
\label{eq:polygon_density}
\end{align}
where $\delta$ is the Dirac delta function, $s_i$ is the arclength of corner $i$, and $\beta_i=\psi_i-\psi_{i-1}$
is the size of the angular sector lying between the outward normals of the two edges meeting at corner $i$ (see Fig.~\ref{fig:sketches}).
Edges occupy a set of measure zero in the space of orientations and thus have zero population. Corner $i$ traps every particle whose orientation $\theta$ lies in the interval $[\psi_{i-1},\psi_i]$ and its  population is proportional to its size $\beta_i$.

The simulation results are shown in Fig.~\ref{fig:polygon_rho}.
As expected, there are sharp density peaks at each corner (top panel),
and the corner populations (the areas under these peaks) are proportional to $\beta$ (bottom left panel).
The fraction of particles not in the vicinity of any corner (the ``edge fraction''), on the other hand, decreases slowly with $L\Dr/v_0$ (bottom right panel), remaining of order $10\%$ at $L\Dr/v_0\sim10^{-3}$.
In that sense, the meaning of ``strong confinement'' is more restrictive for polygonal boxes than for rounder ones. However, Eq.~\eqref{eq:polygon_density} accurately describes the relative corner populations even when edge fractions are large.


\smallskip	
\emph{Conclusion} ---
In summary, we have shown how to predict the density and pressure distribution of a simple active fluid from the geometry of its confining box, provided the box is convex and small enough.
Conversely, our theory predicts the box shape that will yield any desired density profile on the boundary, thus offering the first general tool to understand and design such confinements.

The result relies on the ability of particles to
circumnavigate their container
faster than they re-orient (the strong confinement limit).
While this limit is easily achieved when curvature is positive and misaligned particles move ballistically, our simulation results show that it even can be realized in the extreme case of a flat edge, where particles experience randomly accelerated motion. Thus, the predicted density distribution (Eq.~\eqref{eq:density_variance}) holds for small convex boxes of arbitrary shape.
\\

\begin{acknowledgments}
This research was supported by
NSF-MRSEC-0820492 and NSF-DMR-1149266.
Computational resources were provided by the NSF through XSEDE computing resources and the
Brandeis HPCC.
\end{acknowledgments}


\appendix

\section{Overdamped Approach}
\label{ap:overdamped}

Here we present a derivation based on stochastic calculus of the main result of the paper: the relationship between the density on the boundary and the local curvature radius of the boundary. \\

We start from the linearized equations of motion on the boundary Eqs.~(2) (main text), written in a form that emphasizes their equivalence with a Langevin equation for a free particle with position-dependent friction and temperature:
\begin{align}
v = \dot{s}
\, , \quad
\dot{v} = -\frac{v_0}{R}\,v + v_0\xi(t)
\, ,
\end{align}
where $v=v_0\phi$ and $\langle\xi(t)\xi(t')\rangle=2\Dr\delta(t-t')$.
Then, eliminating the momentum variable $v$ (or $\phi$) by averaging over the fast time scale $R/v_0$ is equivalent to taking the overdamped limit.
Since the friction coefficient is position-dependant, care must be taken to circumvent the ``Ito-Stratonovitch dilemma''.
The problem was solved by Sancho et al.~\cite{Sancho1982}, who find that $s$ obeys the following Stratonovitch stochastic differential equation:
\begin{align}
\dot{s} = R(s) \xi(t)
\, .
\end{align}
The corresponding Smoluchowski equation for the density $\rho(x,t)\equiv\langle\delta(x-s(t))\rangle$ is
\begin{align}
\label{eq:stochastic}
\partial_t \rho = \Dr\partial_x[R\,\partial_x(R\rho)]
\, ,
\end{align}
whose steady-state solution is given by
\begin{align}
\rho (x) = \frac{c_1}{R(x)} \left[ 1 + \int_0^x \frac{c_1 dy}{R(y)} \right]
\label{eq:rhostst}.
\end{align}
The two integration constants $c_1$ and $c_2$ are determined by enforcing the periodicity of $\rho(x)$:
\begin{align}
\rho (x) = \rho(x+L)
\ \implies \
\int_0^L \frac{c_1 dy}{R(y)} = 0
\ \implies \
c_1=0
\end{align}
and the normalization of density:
\begin{align}
\int_0^L dx \rho (x) = N
\ \implies \
\int_0^L \frac{c_0 dx}{R(x)} = N
\ \implies \
c_0 = \frac{N}{2\pi}
\end{align}
where we have used the relation $\oint_0^L dx/R(x)=2\pi$ for plane curves.
The resulting expression for $\rho$:
\begin{align}
\rho (s) = \frac{N}{2\pi R}
\ ,
\end{align}
is identical to that of Eqs.~(6), obtained by moment expansion.

\section{Angular distribution}
\label{ap:var_skew_kurt}

\begin{figure}[h]
\centering
\includegraphics[width=\linewidth]{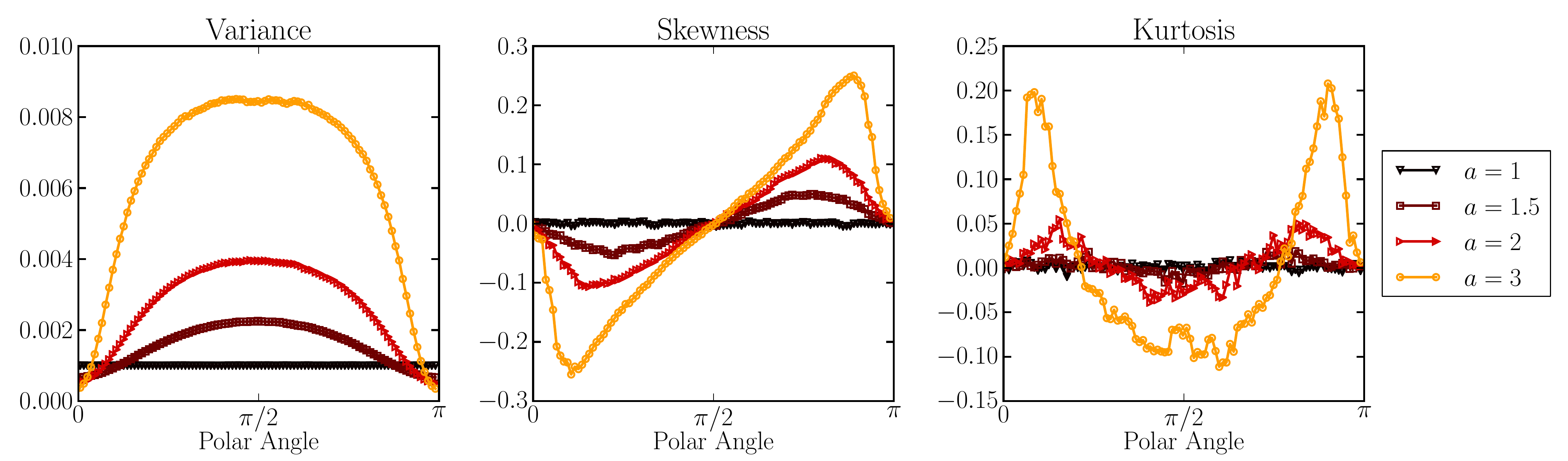} 
\caption{
Variance, skewness and kurtosis of the orientation $\phi$ relative to the boundary normal in small elliptic boxes of aspect ratio $a$.}
\label{fig:var_skw_kurt}
\end{figure}

To assess the importance of higher order moments in Eq.~(6), we show on Fig.~\ref{fig:var_skw_kurt} the variance, skewness and kurtosis of the distribution of the angle $\phi$ between the boundary normal and the particle's orientation in ellipses of various aspect ratios $a$. The data comes from the same runs used for Fig.~2.
As suggested by the mismatch between the prediction $\langle\phi^2\rangle=R\Dr/v_0$ and the numerical observations on Fig.~2 for values of $a$ not close to $1$, higher order moments can only be neglected in almost circular boxes.

\onecolumngrid

\twocolumngrid

\bibliography{confined_active}

\end{document}